# Perineural Invasion Detection in Multiple Organ Cancer Based on Deep Convolutional Neural Network

Ramin Nateghi[1*], Fattaneh Pourakpour[2]

[1]Electrical and Electronics Engineering Department, Shiraz University of Technology, Shiraz, Iran

Iranian Brain Mapping Lab, National Brain Mapping Laboratory, Tehran, Iran

[1*]r.nateghi@sutech.ac.ir, [2] pourakpour@nbml.ir

**Abstract**

Perineural invasion (PNI) by malignant tumor cells has been reported as an independent indicator of poor prognosis in various cancers. Assessment of PNI in small nerves on glass slides is a labor-intensive task. In this study, we propose an algorithm to detect the perineural invasions in colon, prostate, and pancreas cancers based on a convolutional neural network (CNN).

## 1. Introduction

The identification of perineural invasion (PNI) has been proven to be a sign of poor prognosis of patients [1]. The perineural invasion was defined for the first time by Batsakis in 1985 [2]. It is a form of cancer progression in which tumor cells touch or invade nerves. Cancer in patients with PNI is most likely to be spread in the other parts of the body through at least one nerve, and this is the reason for the poor prognosis of these patients. Fig. 1 shows three nerves that have been invaded by tumor cells in which the intersection of nerve tissue with tumor cells has been shown with red lines.

PNI can happen in various cancers, including prostate, pancreatic ductal adenocarcinoma, gastric carcinoma, colorectal and colon, pancreatobiliary tract [2]. Identifying the mechanism of PNI is significantly important for blocking cancer progression and improving patient survival. The prognostic significance of PNI has been evaluated in various researches [3,4]. Zhang et al. evaluated the association between PNI and biochemical recurrence (BCR) in patients with prostate cancer (PCa) [3]. The prognostic role of perineural invasion in surgically treated esophageal squamous cell carcinoma has been investigated by Kim et al [4]. The impact of PNI on the 5-years overall survival of Rectal cancer (RC) patients has been evaluated in [5].

The manual identification of PNI is a time-consuming and labor-intensive task [6]. As mentioned earlier, perineural invasion is composed of nerve and tumor cells attached to the nerve. The morphology and growth properties of tumor cells can be completely different according to the histologic type and organs they are present in, making the identification of PNI even more difficult. In this research, we present a deep learning system to identify PNI in colon, prostate, and pancreatobiliary tract cancers biopsies.

## 2. Material and Methods

For our work, we used the publicly available dataset for the international MICCAI-PAIP 2021 challenge, including 150 training whole slide images (WSI) from three different colon, prostate, and pancreatobiliary tract cancers, with 50 WSIs per cancer. All images were stained with standard hematoxylin and eosin stain. The validation and testing sets consist of 30 and 60 WSIs. The validation set is used for evaluating the proposed method. The testing set is also used for the final ranking on the test leaderboard. The perineural invasion boundaries between nerve tissue and tumor cells are provided as 1-pixel thick lines for training and validation sets. Fig. 2, represents a sample region of a histopathology image and corresponding PNI boundaries as a binary mask.

We developed a convolutional neural network (CNN) based method to precisely identify PNI in histopathology images.

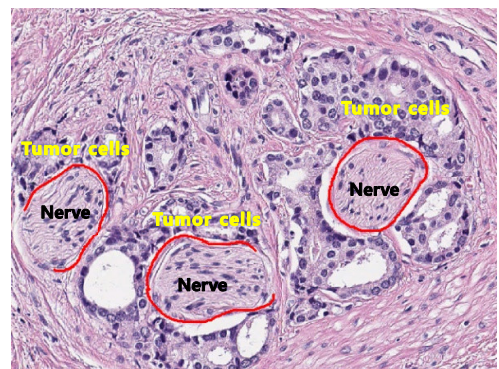

**Fig. 1**: Perineural invasion boundaries

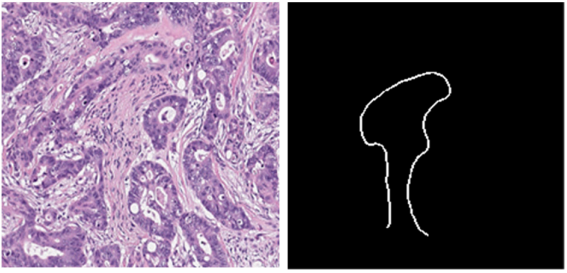

**Fig. 2**: Sample image and corresponding PNI boundary as a binary mask

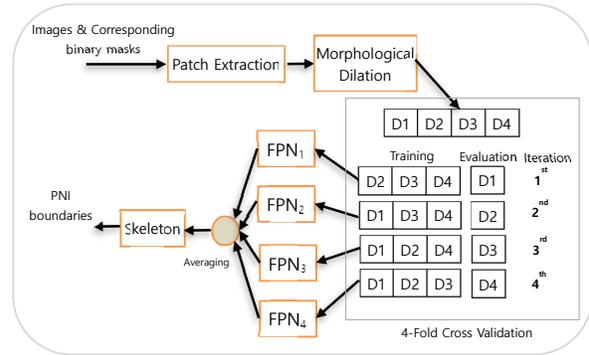

**Fig. 3**: Pipeline of our PNI segmentation method

The entire pipeline of our method is shown in Fig. 3. In the first step, the training images are partitioned to the non-overlapping patches with a size of 512×512. Before training, we applied a morphological dilation operator with a disk-shape structure element (radius 3) to the binary masks to thicken PNI boundaries, helping our model to train more efficiently. We also used various data augmentation techniques, like rotation, translation, scaling, brightness, RGB shift and smoothing transforms to enlarge our training set and improve model generalization as well. All of the patches and corresponding masks are used to train Feature Pyramid Network (FPN) models. FPNs are one of the most popular deep learning-based models for image segmentation which aggregate multi-scale feature information to get more precise segmentation results. For model training, instead of splitting our training data dataset into two training and validation subsets, we used the k-fold cross-validation technique (with k=4) as a preventative technique against overfitting in which the training dataset is split into four different subsets. Each subset is considered as evaluation data once and the remaining sets as training data, through which four FPN models are trained. We also used EfficientNet0 as the base model for FPNs. To train models, we used a mini-batch size of 8, a cyclical maximal learning rate of 10-4 for 60 epochs. The binary cross-entropy+dice loss is also considered to train the models.

During testing, the test image is split into overlapping patches of 512×512 with a stride size of (256,256). Using overlapped patches in the testing phase can effectively reduce the edge artifacts at patch borders and help to get continuous PNI boundaries and more accurate results. The extracted patches are then pass through the FPN models and the results are combined (using averaging) to get the segmentation result. Finally, the morphological skeleton method is used to process our predicted lines as 1-pixel thick lines.

To improve PNI segmentation performance, we also used Test-Time Data Augmentation (TTA) technique by applying some augmentation transforms (rotation, brightness, and contrast, RGB shift, smoothing, and mirroring) on images during inference.

## 3. Results

We evaluated the performance of our method on the validation set. Table .1 summarizes the performance of our method based on the F1-score criterion. When no augmentation was used, the performance was vastly reduced to an F1-score of 25.85%. On the other hand, the overlapping patch extraction technique considerably improved the performance of our method to an F1-score of 38.69%. The splitting WSIs into small patches that was done in the first step may lead to undesirable artifacts and make discontinuous PNI boundaries, especially on the border of extracted patches. We found that the overlapping patch extraction technique can address this problem during inference and provide continuous PNI boundaries. Our best result on the validation set was obtained when using the TTA, overlapping patch extraction, and augmentation techniques, resulting in an F1-score of 41.55%. The TTA technique also improved the performance of our method by aggregating the predictions from different transformations for a given test image. Fig. 4 represents the visual examples of PNI detection by our method.

### Acknowledgment

De-identified pathology images and annotations used in this research were prepared and provided by the Seoul National University Hospital by a grant of the Korea Health Technology R&D Project through the Korea Health Industry Development Institute (KHIDI), funded by the Ministry of Health & Welfare, Republic of Korea (grant number: HI18C0316). The authors thank all organizers and contributors of the MICCAI-PAIP 2021 challenge.

Table 1. Performance of our method on the validation set

| Method | F1-score |
|---|---|
| Without Augmentation | 25.85 % |
| Augmentation + TTA | 30.32 % |
| Augmentation + overlapping patch extraction | 38.69 % |
| Augmentation + overlapping patch extraction + TTA | **41.55 %** |

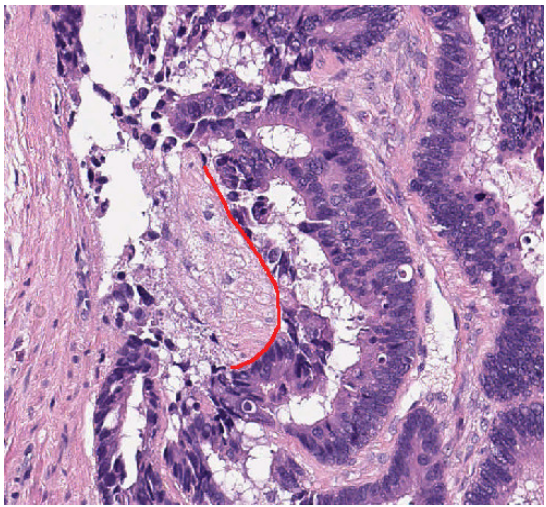
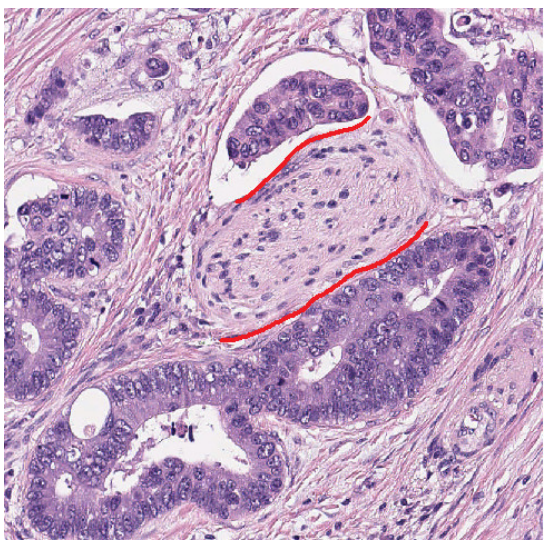
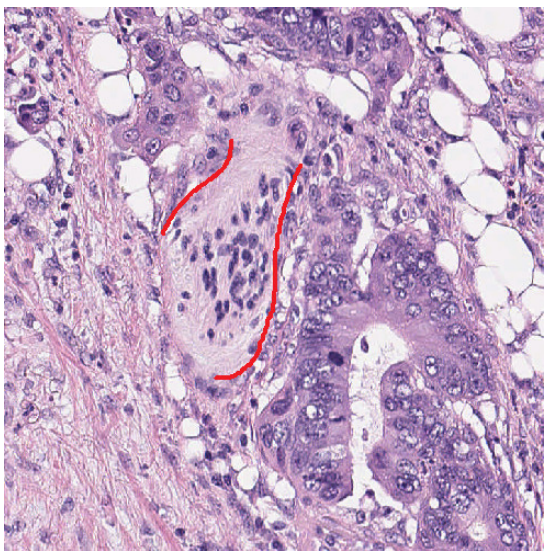

**Fig. 4**: Visual examples of PNI detection by our method